\documentclass[12pt]{article}

\usepackage{amssymb}
\usepackage{subfigure}
\usepackage{color}
\usepackage{epsfig,amssymb,amsfonts,amsmath,graphicx,dsfont,cite,xfrac}
\usepackage{authblk}
\definecolor{mygray}{gray}{0.5}

\usepackage{cite}
\usepackage[colorlinks=true,linkcolor=blue,citecolor=red]{hyperref}

\usepackage{tikz,pgfplots}
\usetikzlibrary{positioning}
\usetikzlibrary{matrix}

\newcommand{\be}{\begin{equation}}
\newcommand{\ee}{\end{equation}}
\newcommand{\bea}{\begin{eqnarray}}
\newcommand{\eea}{\end{eqnarray}}

\title{Quantum nonstationary oscillators: Invariants, dynamical algebras and coherent states via point transformations}

\author[${1,2}$]{K. Zelaya\thanks{zelayame@crm.umontreal.ca}}
\author[${2}$]{Oscar Rosas-Ortiz\thanks{Corresponding author: orosas@fis.cinvestav.mx}}

\affil[${1}$]{\footnotesize Centre de Recherches Math\'ematiques, Universit\'e de Montr\'eal, Montr\'eal, Qu\'ebec H3C 3J7, Canada}

\affil[${2}$]{\footnotesize Physics Department, Cinvestav, AP 14-740, 07000
M\'exico City, Mexico}

\date{}
\begin{document}

\maketitle

\begin{abstract}
We consider the relations between nonstationary quantum oscillators and their stationary counterpart in view of their applicability to study particles in electromagnetic traps. We develop a consistent model of quantum oscillators with time-dependent frequencies that are subjected to the action of a time-dependent driving force, and have a time-dependent zero point energy. Our approach uses the method of point transformations to construct the physical solutions of the parametric oscillator as mere deformations of the well known solutions of the stationary oscillator. In this form, the determination of the quantum integrals of motion is automatically achieved as a natural consequence of the transformation, without necessity of any ans\"atz. It yields the mechanism to construct an orthonormal basis for the nonstationary oscillators, so arbitrary superpositions of orthogonal states are available to obtain the corresponding coherent states. We also show that the dynamical algebra of the parametric oscillator is immediately obtained as a deformation of the algebra generated by the conventional boson ladder operators. A number of explicit examples is provided to show the applicability of our approach.

\end{abstract}


\section{Introduction}

The dynamics of many physical systems is described by using quantum time-dependent harmonic oscillators \cite{Dod75,Pri83,Cum86,Cum88,Pro91,Ghe92,Dod95,Dod05,Maj05,Mih09,Cor11,Der13,Gue15,Leo16,Zha16,Zel17a,Con17,HCr18,Con19}, where the construction of minimum wave packets is relevant \cite{Har82,Com12,Cas13,Sch13,Cru15,Cru16,Afs16,Mih18,Mih19,Una18,Zel19} (see also the recent reviews \cite{Dod18,Ros19}). Such a diversity of applications is due to the quadratic profile of the oscillator \cite{Dod95,Man96,Dod00a,Dod00b,Cor10,Nag19,Ram18,Wol81,Dod89}, which is also useful in the trapping of quantum particles with electromagnetic fields \cite{Pri83,Cum86,Ghe92,Maj05,Mih09,Mih18,Mih19,Pau90,Gla92,Bar96,Dod96,Dod98,Cas98,Gen11,Cas12}. In most of the cases reported in the literature the oscillator has a frequency of oscillation that depends on time. Usually, it is also acted by a driving force which also depends on time. Thereby, the oscillator is subjected to external forces that either take energy from it or supply energy to it. Such a nonconservative system has no solutions with the property of being orthogonal if they are evaluated at different times. Nevertheless, diverse techniques have been developed to find solutions with physical meaning \cite{Wol81,Dod89,Dod95,Gla92,Dod00a,Dod00b,Cor10,Cas13,Sch13,Cru15,Cru16,Ram18,Nag19}. The progenitor of most of the solvable models reported in the literature is the approach of Lewis and Reisenfeld \cite{Lew68,Lew69}, where an invariant operator is introduced, as an ans\"atz,  to get a basis of eigenvectors that serve to construct the physical solutions. Important results on the matter were obtained by Dodonov and Man'ko \cite{Dod89}, and by Glauber \cite{Gla92}. Further developments have been reported in, e.g. \cite{Dod95,Zha16,Cru15,Cru16,Dod00a,Dod00b,Cor10,Nag19}. 

In the present work we develop an approach to study nonstationary oscillators by means of the so called point transformations \cite{Dew52,Ste93}. These have been used in the classical context to deform the trajectories of a given linear second order differential equation into trajectories of the free particle \cite{Arn83}, although the latter procedure is commonly called {\em Arnold transformation}. An extension to quantum systems was introduced in \cite{Ald11} which, in turn, has been used to study the Caldirola-Kanai oscillator \cite{Gue12,Gue13} (see also the book \cite{Sch18}). The point transformations are also useful to interrelate the harmonic oscillator with a series of oscillator-like systems for which the mass is a function of the position \cite{Cru09,Cru13}, as well as to study the ordering ambiguity of the momentum operator for position-dependent mass systems in the quantum case \cite{Mus19}. The major advantage of the point transformation method is that conserved quantities (first integrals) as well as the structure of the inner product are preserved \cite{Ste93}. Another property of these transformations is that they can be constructed to be invertible. Then, one may depart from a system, for which the dynamical law of motion is already solved, to arrive at a new exactly solvable dynamical law that can be tailored on demand to describe the behavior of another system, and vice versa.

In the present case we are interested in solving the Schr\"odinger equation associated to the Hamiltonian
\be
\hat{H} (t)=\frac{\hat{p}^{2}}{2m}+\frac{m}{2}\Omega^{2}(t)\hat{x}^{2}+F(t)\hat{x}+V_{0}(t) \mathbb{I},
\label{eq:PMO1}
\ee
where $\hat x$ and $\hat p$ are the canonical operators of position and momentum  $[\hat{x},\hat{p}]=i\hbar \mathbb{I}$, $F(t)$ stands for a time-dependent driving force, $V_{0}(t)$ is the time-dependent zero point energy, and $\mathbb I$ is the identity operator. The function $\Omega(t)$ is real-valued and positive. That is, the Hamiltonian (\ref{eq:PMO1}) describes a nonstationary oscillator,  the frequency of which $\Omega(t)$ depends on time. In general, the system under interest is nonconservative, so the orthogonality of the related solutions is not granted a priori. As $\hat H$ is not an integral of motion, an additional problem is to determine the invariants (first integrals) that may serve as observables to define uniquely the system. 

The main result reported in this work is to show that the properly chosen point transformations permit to solve the above problems by overpassing the difficulties that arise in the conventional approaches. In particular, we show that the integrals of motion are automatically obtained as a consequence of the transformation, without necessity of any ans\"atz. Another interesting result is that the point transformations permit to verify the orthogonality of the basis states, so that the construction of arbitrary linear superpositions is achieved easily. The latter lays the groundwork to construct the corresponding coherent states since the dynamical algebras are also immediately obtained as a deformation of the well known boson algebra.

The paper is organized as follows. In Section~\ref{oscilador} we pose the problem to solve by providing the explicit forms of the Schr\"odinger equation for the stationary oscillator and the nonstationary one. In Section~\ref{point} we solve the differential equation of the parametric oscillator by point transforming the differential equation of the stationary one. In Section~\ref{ortogonal} we verify that the orthogonality of the initial solutions as well as the matrix representation of observables is inherited to the new system by the point transformations. The determination of the invariants (quantum integrals of motion) for the new system is discussed in Section~\ref{integrals}, and the derivation of the related dynamical algebras is developed in Section~\ref{dynamical}. We discuss the superposition of the solutions of the nonstationary oscillators in Section~\ref{Seclin}. The construction of the coherent states of the parametric oscillator is developed in Section~\ref{Seccs}, where we show that these states share almost all the properties of the Glauber states \cite{Gla07}, except in the fact that they minimize the Schr\"odinger-Robertson inequality rather than the Heisenberg uncertainty. Section~\ref{examples} provides some particular cases as concrete examples of the applicability of our approach. Some results reported already by other authors are recovered on the way. Final concluding remarks are given in Section~\ref{conclu}. Detailed information about the point transformations we use throughout the manuscript is provided in Appendix~\ref{ApA}. A discussion about the possibility of making the zero point energy $V_0(t)$ equal to zero without loosing generality is delivered in Appendix~\ref{ApB}. Finally, relevant information about the Ermakov equation, which is a keystone in our approach, can be found in Appendix~\ref{ApC}.

\section{One-dimensional parametric oscillator}
\label{oscilador}

The one-dimensional stationary quantum oscillator with mass $m$ and constant frequency of oscillation $w$ is described by the Hermitian Hamiltonian
\be
\hat{H}_{osc}=\frac{\hat{P}^{2}}{2m}+\frac{m}{2}w^{2}\hat{X}^{2}, \quad w>0,
\label{eq:INT0}
\ee
where $\hat{X}$ and $\hat{P}$ stand for the canonical position and momentum operators, $[\hat{X},\hat{P}] = i \hbar$. The Schr\"odinger equation for the oscillator wave function $\Psi(X,\tau)=\langle X \vert \Psi(\tau)\rangle$ in the position representation is well known
\be
i\hbar\frac{\partial\Psi}{\partial \tau} = - \frac{\hbar^2}{2m}\frac{\partial^{2}\Psi}{\partial X^2} + \frac{1}{2}m w^2 X^2\Psi=0,
\label{eq:INT1}
\ee
with $\tau$ the time-parameter. The solutions are easily achievable by separation of variables $\Psi(X,\tau)=e^{-i E \tau/\hbar}\Phi(X)$, where $\Phi(X)= \langle X\vert\Phi \rangle$ fulfills the eigenvalue equation
\be
-\frac{\hbar^{2}}{2m}\frac{d^{2}\Phi}{dX^2} + \frac{1}{2}mw^{2}X^2 \Phi = E \Phi.
\label{eq:INT2-2}
\ee
The fundamental set of normalized solutions is therefore
\be
\Phi_{n}(X)=\sqrt{\frac{1}{2^{n}n!}\sqrt{\frac{mw}{\pi\hbar}}} \, e^{-\frac{mw}{2\hbar}X^{2}}H_{n}\left(\sqrt{\frac{mw}{\hbar}}X
\right), \quad E_{n}=\hbar w(n+1/2),
\label{eq:INT3}
\ee
where $H_n(z)$ are the Hermite Polynomials \cite{Olv10}. In the space ${\cal H} = \mbox{span} \{\vert\Phi_{n}\rangle\}_{n=0}^{\infty}$, a vector $\vert \Phi \rangle$ is regular if it satisfies the normalization condition $\vert\vert \vert\Phi\rangle\vert\vert^{2}=\langle \Phi\vert\Phi\rangle<\infty$, with  inner product defined as follows
\be
\langle \Phi_{(2)}\vert\Phi_{(1)}\rangle=\int_{-\infty}^{\infty}dX\,\Phi_{(2)}^{*}(X)\Phi_{(1)}(X) \, .
\label{eq:INT2-3}
\ee
Clearly, the basis set is orthonormal $\langle\Phi_n \vert\Phi_m \rangle=\delta_{n,m}$. 

On the other hand, the wave functions $\psi(x,t)=\langle x \vert \psi(t)\rangle$ of the one-dimensional non stationary quantum oscillator described by the Hamiltonian \eqref{eq:PMO1} satisfy the Schr\"odinger equation
\be
i\hbar\frac{\partial\psi}{\partial t} = -\frac{\hbar^{2}}{2m}\frac{\partial^2\psi}{\partial x^2} \psi + \frac{1}{2}m\Omega^{2}(t)x^2 \psi +F(t)x \psi +V_{0}(t) \psi .
\label{eq:INT4}
\ee
In this case the oscillator has a frequency of oscillation $\Omega$ that depends on time. The driving force $F$ and zero point of energy $V_0$ also depend on time. That is, the oscillator under study is subjected to external forces that either take energy from it or supply energy to it. This system is nonconservative, with no orthogonal basis of solutions $\psi_n(x,t)$ at arbitrary times $t$ and $t'$, $\langle \psi_n(t) \vert \psi_m (t') \rangle \neq \delta_{n,m}$ for $t\neq t'$. Nevertheless, as it has been indicated in the introduction, diverse techniques have been developed to find solutions with physical meaning \cite{Wol81,Dod89,Dod95,Lew68,Lew69,Gla92,Dod00a,Dod00b,Cor10,Cas13,Sch13,Cru15,Cru16,Ram18,Nag19}. 

In the sequel we show that the Schr\"odinger equations (\ref{eq:INT1}) and (\ref{eq:INT4}) are interrelated in such a form that the solutions of the stationary problem (\ref{eq:INT1}) can be used to get the solutions of the nonstationary one (\ref{eq:INT4}), and vice versa. The key is provided by a deformation of the coordinate variable, the time parameter, and the wave functions of the `initial' system, which gives rise to the corresponding variables and parameters of the `new' (or `deformed') system. Such a deformation is properly defined by point transformations \cite{Ste93}. We shall consider the stationary oscillator as the initial system, so the parametric oscillator can be interpreted as a deformation of the stationary one.

\subsection{Point transformations}
\label{point}

We look for relationships between the elements of the set $\{ X, \tau, \Psi \}$ and those of the set $\{ x, t, \psi \}$. Formally,
\be
X=X(x,t), \quad  \tau =  \tau (x,t), \quad \Psi = \Psi (X(x,t), \tau (x,t)).
\label{eq:INT5}
\ee
Notice that the dependence of $\Psi$ on $x$ and $t$ is implicit, so it is convenient to rewrite it as an explicit function of the elements in $\{ x, t, \psi \}$. We may write
\be
\Psi = G(x,t;\psi(x,t)).
\label{eq:INT5-1}
\ee
The explicit dependence of $G$ on $\psi$ is essential, since it provides a mechanism to map any solution of~\eqref{eq:INT1} into the set of solutions of~\eqref{eq:INT4}, and vice versa. To be precise, the latter equations are respectively of the form
\be
S_{in}  \left(X, \tau; \Psi, \Psi_{\tau},  \Psi_{X,X} \right) =0,  \quad S_{def} \left( x,t; \psi, \psi_t, \psi_{x,x} \right)=0,
\label{eq:INT6-1}
\ee
with nonlinearities present in neither $S_{in}$ nor $S_{def}$. Hereafter, for simplicity,  we use no-number subindices to denote partial derivatives $f_u = \frac{\partial f}{\partial u}$.

Departing from $S_{in}$, the proper point transformation (see Appendix~\ref{ApA} for details) produces
\be
i\hbar \psi_t+\frac{\hbar^{2}}{2m}\frac{\tau_{t}}{X_{x}^{2}} \psi_{x,x} + B(x,t) \psi_x - V(x,t)\psi = 0,
\label{eq:INT11}
\ee
where
\be
\begin{aligned}
& B(x,t)=-i\hbar\frac{X_{t}}{X_{x}}+\frac{\hbar^{2}}{2m}\frac{\tau_{t}}{X_{x}^{2}}\left( 2\frac{A_{x}}{A}-\frac{X_{xx}}{X_{x}} \right) ,\\[1ex]
& V(x,t)=-i\hbar\left(\frac{A_{t}}{A}-\frac
{X_{t}}{X_{x}}\frac{A_{x}}{A} \right)-\frac{\hbar^{2}}{2m}\frac{\tau_{t}}{X_{x}^{2}}\left( \frac{A_{xx}}{A}-\frac{X_{xx}}{X_{x}}\frac{A_{x}}{A} \right)+\frac{\tau_{t}}{2}m w^{2}X^{2}(x,t).
\end{aligned}
\label{eq:INT12}
\ee
As Eq.~(\ref{eq:INT11}) must be of the form $S_{def}$ indicated in (\ref{eq:INT6-1}), we impose the conditions
\be
\frac{\tau_{t}}{X_{x}^{2}}=1, \quad B(x,t)=0.
\label{eq:INT13}
\ee
To satisfy the first condition let us introduce a real-valued function $\sigma(t) >0$ such that $\tau_{t}=\sigma^{-2}(t)$. Then, by simple integration (and some rearrangements), one gets 
\be
\tau (t)=\int^{t}\frac{dt'}{\sigma^{2}(t')}, \quad X(x,t)=\frac{x+\gamma(t)}{\sigma(t)},
\label{eq:INT14}
\ee
where the real-valued function $\gamma(t)$ stems from the integration with respect to $x$. Clearly $X_{xx}=0$ for any functions $\sigma >0$ and $\gamma$. Then, the condition $B(x,t)=0$ leads to
\be
A(x,t)=\exp\left[ i\frac{m}{\hbar}\left(-\frac{\dot{\sigma}}{2\sigma}x^{2}+\frac{W}{\sigma}x+\eta\right)\right], \quad W(t)=\sigma\dot{\gamma}-\dot{\sigma}\gamma, 
\label{eq:INT15}
\ee
with $\dot f = \frac{df}{dt}$, and $\eta=\eta(t)$ a complex-valued function that arises by integration. The introduction of \eqref{eq:INT15} into \eqref{eq:INT12} gives the energy potential
\bea
V(x,t) = \frac{m}{2}\left(-\frac{\ddot{\sigma}}{\sigma}+\frac{w^{2}}{\sigma^{4}}\right)x^{2}+m\left(\frac{\dot{W}}{\sigma}+w^{2}\frac{\gamma}{\sigma^{4}}\right)x 
+\frac{m}{2}\left( i\frac{\hbar}{m}\frac{\dot{\sigma}}{\sigma}+2\dot{\eta}-\frac{W^{2}}{\sigma^{2}}+w^{2}\frac{\gamma^{2}}{\sigma^{4}}\right).
\label{eq:INT16}
\eea
Comparing this result with Eq.~(\ref{eq:INT4}) we obtain a system of three equations for $\sigma$, $\gamma$, and $\eta$. Without loss of generality we may take $V_0(t)=0$ (see Appendix~\ref{ApB}) to get
\be
\ddot{\sigma}+\Omega^{2}(t)\sigma=\frac{w^2}{\sigma^{3}}, \quad \ddot{\gamma}+\Omega^{2}(t)\gamma=\frac{F(t)}{m}, \quad \eta(t)=\xi(t)-i\frac{\hbar}{2m}\ln\sigma(t),
\label{eq:INT17}
\ee
where the real-valued function $\xi(t)$ is given by
\be
\xi(t)=\frac{\gamma W}{2\sigma}-\frac{1}{2m}\int^{t}dt'F(t')\gamma(t').
\ee
Remark that $\xi$ is just a displaced version of $\eta$ in the complex plane that permits to rewrite the function $A(x,t)$ in (\ref{eq:INT15}) as follows
\be
A(x,t)=\sqrt{\sigma}\exp\left[ i\frac{m}{\hbar}\left(-\frac{\dot{\sigma}}{2\sigma}x^{2}+\frac{W}{\sigma}x+\xi\right)\right].
\label{eq:INT18}
\ee
In turn, the time-dependent function $\sigma$ satisfies the Ermakov equation~\cite{Erm08}, which is a quite natural result in the studies of the parametric oscillator \cite{Cas13,Sch13,Cru15,Cru16}. Therefore, for a set of nonnegative parameters $\{a,b,c\}$, we have
\be
\sigma(t)= \left[ a q_1^2(t)+ b q_1(t)q_2(t)+c q_2^2(t) \right]^{1/2},
\label{eq:OSC7}
\ee
where $q_{1}$ and $q_{2}$ are two linearly independent real solutions of the linear homogeneous equation obtained from (\ref{eq:INT17}) by making $w=0$, see Appendix~{\ref{ApC} for details. That is, the Wronskian $W(q_1,q_2) =W_0$ is a constant. The condition $b^2-4ac=- 4\tfrac{w^2}{W_{0}^2}$ ensures $\sigma >0$ at any time\cite{Ros15,Bla18}. Notice that $w\rightarrow 0$ produces $b=2 \sqrt{ac}$, so that $\sigma_{free} = \sqrt{a} q_1 + \sqrt  c q_2$. That is, our method applies even if the initial Hamiltonian $\hat H_{osc}$ in (\ref{eq:INT0}) is reduced to the purely kinematic Hamiltonian of the free particle. The deformation of the system is thus provided by the point transformation ruled by the function $\sigma_{free}$, although the latter is not necessarily connected with the parametric oscillator. In the present work we omit the analysis of such a case, results on the matter will be reported elsewhere.

On the other hand, $\gamma(t)$ describes a classical oscillator of frequency $\Omega(t)$ that is subjected to the driving force $F(t)$, see e.g. \cite{Ros08}. This function can be expressed as the sum of the homogeneous solution $\gamma_h = \gamma_{1} q_{1}(t) + \gamma_{2} q_{2}(t)$, and an arbitrary particular solution $\gamma_{p}(t)$. The real constants $\gamma_{1,2}$ as well as the function $\gamma_{p}(t)$ are defined whenever the driving force $F(t)$ has been provided. Therefore, the function $\tau$ introduced in (\ref{eq:INT14}) can be rewritten in terms of $q_1$ and $q_2$:
\be
\tau (t)=\int^{t}\frac{dt'}{\sigma^{2}(t')}=\frac{1}{w}\arctan\left[ \frac{W_0}{2w}\left( b+2c\frac{q_2}{q_1} \right) \right].
\label{eq:OSC8}
\ee

To conclude this section we emphasize that, as a result of the point transformation, the function (\ref{eq:INT5-1}) acquires the factorized form $\Psi = G(x,t; \psi (x,t))=A(x,t) \psi(x,t)$, see Appendix~\ref{ApA}. Therefore, we can write the solutions $\psi(x,t)$ of the parametric oscillator in terms of the solutions $\Psi(X,\tau)$ of the stationary one, and vice versa. As we have already solved the stationary case, it is easy to get  the solutions we are looking for
\be
\psi(x,t)=\exp\left[ i\frac{m}{\hbar}\left(\frac{\dot{\sigma}}{2\sigma}x^{2}-\frac{W}{\sigma}x-\xi\right)\right]\frac{\Psi(X(x,t), \tau(t))}{\sqrt{\sigma}} \, .
\label{eq:INN1}
\ee


\subsection{Orthogonality and basic solutions}
\label{ortogonal}

As indicated above, the explicit form of the solutions $\psi_n(x,t)$ is easily achieved from (\ref{eq:INN1}) by using $\Psi_n (X,\tau)=e^{-i E_n \tau/\hbar} \Phi_n (X)$ and the functions $\Phi_n(X)$ defined in (\ref{eq:INT3}). However, the orthogonality of the new set $\psi_n(x,t)$ is not evident. We are interested in the orthogonality of these functions since, although it is not a necessary condition to get physically admissible solutions, it is sufficient to get superpositions of states in easy form. To elucidate such a property let us consider a pair of arbitrary solutions of the stationary oscillator, $\Psi_{(1)}(X, \tau)$ and $\Psi_{(2)}(X, \tau)$. Using (\ref{eq:INN1}), the straightforward calculation gives
\be
\int^{\infty}_{-\infty}dX \, \Psi_{(2)}^{*}(X, \tau)\Psi_{(1)}(X, \tau) = \int^{\infty}_{-\infty}dx \, \psi_{(2)}^{*}(x,t)\psi_{(1)}(x,t).
\label{eq:INN2}
\ee
That is, the point transformation preserves the structure of the inner product. Hence, the orthogonal set of solutions $\{ \vert\Psi_{n} (\tau) \rangle \}_{n=0}^{\infty}$ is mapped to an orthogonal set $\{ \vert\psi_{n}(t)\rangle \}_{n=0}^{\infty}$. In position representation one has
\be
\psi_{n}(x,t)=e^{-i \hbar w(n+1/2) \tau(t)} \varphi_{n}(x,t) ,
\label{eq:INN2-1}
\ee
with
\be
\begin{alignedat}{3}
& \varphi_{n}(x,t)&&=A^{-1}(x,t)\Phi\left(\frac{x+\gamma}{\sigma}\right) \\
& &&=\exp\frac{m}{\hbar}\left[ \left(-\frac{w}{\sigma^2}+i\frac{\dot{\sigma}}{\sigma}\right)\frac{x^2}{2} - \left(w\frac{\gamma}{\sigma^2}+i\frac{W}{\sigma}\right)x+\left(-\frac{w}{2}\frac{\gamma^2}{\sigma^2}-i\xi \right) \right] \\
& && \hspace{40mm}\times \sqrt{\frac{1}{2^{n}n!}\sqrt{\frac{mw}{\pi\hbar}}} \frac{1}{\sqrt{\sigma}} H_{n}\left[\sqrt{\frac{mw}{\hbar}}\left(\frac{x+\gamma}{\sigma} \right) \right] \, .
\end{alignedat}
\label{eq:INN3}
\ee
The above expression is in agreement with the results reported by Glauber \cite{Gla92}. From (\ref{eq:INN2}) we immediately realize that the orthonormality 
\be
\int^{\infty}_{-\infty}\, dX \, \Psi_{n}(X, \tau)\Psi^{*}_{m}(X, \tau)=\int^{\infty}_{-\infty}\, dx \, \psi_{n}(x,t)\psi^{*}_{m}(x,t)=\delta_{n,m}
\label{eq:INN4}
\ee
holds when the functions $\psi$ are evaluated at the same time. In general, if $t \neq t'$, the orthonormality is not granted. We write
\be
\int_{-\infty}^{\infty}dx \,\psi_{n}(x,t)\psi_{m}^{*}(x,t') \not=\delta_{n,m}, \quad t\not=t'.
\label{eq:INN5}
\ee
Having in mind that the products (\ref{eq:INN3}) are evaluated at a given time $t$, we may write $\mathcal{H}(t)=\operatorname{Span}\{\vert\psi_{n}(t)\rangle \}_{n=0}^{\infty}$. That is, the space of states we are dealing with is dynamical (see, e.g. \cite{Ali18} for a discussion on the matter). The detailed analysis of the properties of such a space is out of the scope of the present work, so it will be provided elsewhere.

\subsection{Quantum integrals of motion}
\label{integrals}

The nonconservative system described by the Hamiltonian $\hat H(t)$ defined in (\ref{eq:PMO1}), equivalently by the Schr\"odinger equation (\ref{eq:INT4}), is quite different from the stationary oscillator associated to the well known Hamiltonian $\hat H_{osc}$ of Eq.~(\ref{eq:INT0}). Although we have shown the orthonormality of the solutions $\psi_n(x,t)$, it is necessary to emphasize that they are not eigenfunctions of the Hamiltonian $\hat H(t)$. Indeed, the time-dependence of $\hat H(t)$ prohibits the factorization of $\psi(x,t)$ as the product of a purely time-dependent function $T (t)$ with a position-dependent function $\chi (x)$, where $\chi (x)$ fulfills a given eigenvalue equation. Nevertheless, the functions $\psi_n(x,t)$ are admissible from the physical point of view. Since $\hat H(t)$ is not a constant of motion of the system $\frac{d}{dt}\hat{H}(t)\not=0$, we wonder about the observable(s) that define the system uniquely. Such observable(s) must include the set $\psi_n(x,t)$ as its (their) eigenfunctions. Moreover, what about the related spectrum? The latter points must be clarified in order to provide the functions (\ref{eq:INN2-1}), and any linear combination of them, with a physical meaning.

Remarkably, such information is obtained from the point transformation itself, because any conserved quantity is preserved \cite{Ste93}. Indeed, from \eqref{eq:INT3} we see that the energy eigenvalues $E_{n}=\hbar w(n+1/2)$ of the stationary oscillator must be preserved since they are constant quantities. To be specific, using the relationships~\eqref{eq:INT10} of Appendix~\ref{ApA},  the stationary eigenvalue equation~\eqref{eq:INT2-2} gives rise to the new eigenvalue equation
\be
\begin{aligned}
-\sigma^2\frac{\hbar^{2}}{2m}\frac{\partial^2\varphi_{n}}{\partial x^2}&+\frac{m}{2}\left( \dot{\sigma}^{2}+\frac{w^2}{\sigma^2} \right)x^2 \varphi_{n}-\sigma\dot{\sigma}\frac{\hbar}{2i}\left(2x\frac{\partial}{\partial x} + 1 \right)\varphi_{n} + \frac{\hbar\sigma W}{i}\frac{\partial\varphi_{n}}{\partial x} \\ & + m\left(w^2 \frac{\gamma}{\sigma^2}-W\dot{\sigma} \right)x \varphi_{n} +\frac{m}{2}\left(W^{2}+w^2\frac{\gamma^2}{\sigma^2}\right) \varphi_{n} = E_n \varphi_{n},
\end{aligned}
\label{eq:INV1}
\ee
where the eigenvalues $E_{n}=\hbar w(n+1/2)$ have been inherited from the stationary oscillator. It is immediate to identify the operator
\begin{multline}
\hat{I} (t)=\frac{\sigma^2}{2m}\hat{p}^2+\frac{m}{2}\left( \dot{\sigma}^{2}+\frac{w^{2}}{\sigma^2} \right)\hat{x}^2-\frac{\sigma\dot{\sigma}}{2}(\hat{x}\hat{p}+\hat{p}\hat{x})+\sigma W \hat{p} \\ 
+m\left(w^{2}\frac{\gamma}{\sigma^{2}}-W\dot{\sigma} \right)\hat{x} + \frac{m}{2}\left(W^{2}+w^{2}\frac{\gamma^2}{\sigma^2} \right) \mathbb{I}(t), 
\label{eq:INV2}
\end{multline} 
where $\mathbb I(t)$ is the identity operator in ${\cal H}(t)$, see Section~\ref{Seclin}. The operator $\hat I$ is such that the eigenvalue equation 
\be
\hat{I} (t)\vert\varphi_{n}(t) \rangle = \hbar w(n+1/2)\vert\varphi_{n}(t)\rangle
\label{eq:INV1-1}
\ee
coincides with (\ref{eq:INV1}) in position-representation $\varphi_{n}(x,t)=\langle x \vert \varphi_{n}(t)\rangle$. Besides, the straightforward calculation shows that $\hat{I}(t)$ satisfies the invariant condition 
\be
\frac{d}{dt}\hat{I}  (t)=i\hbar[\hat{H}(t),\hat{I} (t)]+\frac{\partial}{\partial t}\hat{I}(t)=0.
\label{eq:INV3}
\ee
That is, $\hat{I}(t)$ is an integral of motion of the parametric oscillator. 

We would like to stress that the invariant operator $\hat{I}(t)$ arises in natural form from the point transformation we are presenting in this work, without necessity of any ans\"atz. In particular, for $\gamma_{1}=\gamma_{2}=F(t)=0$, the operator (\ref{eq:INV2}) coincides with the invariant of Lewis and Reisenfeld~\cite{Lew69}.

\subsection{Dynamical algebra and quadratures}
\label{dynamical}

In addition to the previous results, it is possible to obtain a set of the ladder operators for the parametric oscillator. We first recall that the action of the boson ladder operators
\be
\hat{a}=\frac{\hbar}{\sqrt{2m}}\frac{\partial}{\partial X} + \sqrt{\frac{m}{2}}w X, \quad \hat{a}^{\dagger}=-\frac{\hbar}{\sqrt{2m}}\frac{\partial}{\partial X} + \sqrt{\frac{m}{2}}w X, \quad [\hat{a},\hat{a}^{\dagger}]=\hbar w \mathbb{I}
\label{eq:ALG1}
\ee
on the eigenstates of $H$ is well known 
\be
\hat{a}\Phi_{n+1}(X)=\sqrt{\hbar w (n+1/2)}\Phi_{n}(X), \quad \hat{a}^{\dagger}\Phi_{n}(X)=\sqrt{\hbar w (n+1/2)}\Phi_{n+1}(X).
\label{eq:ALG2}
\ee
The above results are quite natural considering the relationships
\be
\hat{H}_{osc}=\hat{a}^{\dagger}\hat{a}+\frac{\hbar w}{2}, \quad [\hat{H}_{osc},\hat{a}]=-\hbar w \hat{a}, \quad [\hat{H}_{osc},\hat{a}^{\dagger}]=\hbar w \hat{a}^{\dagger}.
\label{eq:ALG1-1}
\ee
Using the relationships \eqref{eq:INT10} of Appendix~\ref{ApA}, the boson operators (\ref{eq:ALG1}) are deformed as follows
\be
\begin{aligned}
& \hat{a}_{2}(t)=\frac{\hbar}{\sqrt{2m}}\sigma\frac{\partial}{\partial x}+\sqrt{\frac{m}{2}}\left( - i \dot{\sigma} +\frac{w}{\sigma} \right)x +\sqrt{\frac{m}{2}} \left(iW+w\frac{\gamma}{\sigma} \right), \\[1ex]
& \hat{a}^{\dagger}_{2}(t)=-\frac{\hbar}{\sqrt{2m}}\sigma\frac{\partial}{\partial x}+\sqrt{\frac{m}{2}}\left( i \dot{\sigma} +\frac{w}{\sigma} \right)x +\sqrt{\frac{m}{2}} \left(-iW+w\frac{\gamma}{\sigma} \right),
\end{aligned}
\label{eq:ALG3}
\ee
while the equations (\ref{eq:ALG2}) acquire the form
\be
\hat{a}_{2}(t) \varphi_{n+1}(x,t)=\sqrt{\hbar w \left(n+\frac{1}{2}\right)} \, \varphi_{n}(x,t), \quad \hat{a}_{2}^{\dagger}(t)\varphi_{n}(x,t)=\sqrt{\hbar w \left( n+\frac{1}{2} \right)}\varphi_{n+1}(x,t).
\label{eq:ALG5}
\ee
Remarkably, the time-dependent ladder operators (\ref{eq:ALG3}}) satisfy the Heisenberg algebra 
\be
[\hat{a}_{2}(t) , \hat{a}_{2}^{\dagger}(t)]=\hbar w \mathbb{I}(t),
\label{algebra}
\ee
and factorize the invariant operator of the parametric oscillator
\be
\hat{I}(t)=\hat{a}_{2}^{\dagger}(t)\hat{a}_{2}(t)+\frac{\hbar w}{2}.
\label{factor}
\ee
The latter leads to the commutation rules
\be
[\hat{I}(t),\hat{a}_{2}(t)]=-\hbar w \hat{a}_{2}(t), \quad [\hat{I}(t),\hat{a}^{\dagger}_{2}(t)]=\hbar w \hat{a}^{\dagger}_{2}(t),
\label{eq:ALG4}
\ee
which verify that $\hat{a}_{2}(t)$ and $\hat{a}^{\dagger}_{2}(t)$ are indeed ladder operators for the eigenfunctions of the invariant operator. On the other hand, the canonical operators of position and momentum become time-dependent 
\be 
\hat{x}=\frac{\sigma}{\sqrt{2m} \, w } \left( \hat{a}_{2}(t)+\hat{a}_{2}^{\dagger}(t) \right)-\gamma \mathbb{I}(t) \, , \quad \hat{p}=\sqrt{\frac{m}{2}}\left( \Xi \, \hat{a}_{2}(t) + \Xi^{*} \, \hat{a}^{\dagger}_{2}(t) \right) - m\dot{\gamma}\mathbb{I}(t),
\label{eq:ALG6}
\ee
where $\Xi(t)=-\frac{i}{\sigma}+\frac{\dot{\sigma}}{w}$. It may be proved that $[\hat{x},\hat{p}]=i\hbar \mathbb{I}(t)$, as expected. 

Using $\hat{I}(t)$, from \eqref{eq:INN2-1}, we find
\begin{subequations}
\begin{equation}
\vert\psi_{n}(t)\rangle=e^{-i\hat{I} (t) \tau(t)/\hbar}\vert\varphi_{n}(t)\rangle,
\label{eq:INV3-1}
\end{equation}
\begin{equation}
\psi_{n}(x,t)=e^{-iw(n+1/2) \tau (t)}\varphi_{n}(x,t).
\label{eq:INV3-2}
\end{equation}
\end{subequations}
Contrary to the stationary case, the operator $e^{-i\hat{I} (t) \tau (t)/\hbar}$ in~\eqref{eq:INV3-1} is not the time evolution operator. No matter it adds the appropriate time-dependent complex phase to the eigenfunctions of $\hat I(t)$, just as this has been discussed by Lewis and Reisenfeld, see Figure~\ref{fig:DIA}.

\begin{figure}
\centering
\begin{tikzpicture}
  \matrix (m) [matrix of math nodes,row sep=5em,column sep=8em,minimum width=2em]
  {
     i\frac{\partial}{\partial \tau}\Psi=\hat{H}_{osc}\Psi
     & i\frac{\partial}{\partial t}\psi=\hat{H}(t)\psi \\
     \hat{H}_{osc}\Phi_{n}=\hbar w(n+1/2)\Phi_{n} 
     & \hat{I}(t)\varphi_{n}=\hbar w(n+1/2)\varphi_{n} \\};
  \path[-stealth]
    (m-1-1) edge [<->,very thick,red] node [left] {$\Psi_{n}=e^{-i\hat{H}_{osc}\tau/\hbar}\Phi_{n}$} (m-2-1)
    (m-2-2) edge [<->,very thick,red] node [right] {$\psi_{n}=e^{-i\hat{I}_{2}\tau(t)/\hbar}\varphi_{n}$} (m-1-2)
   	(m-1-1) edge [blue] node [above] {\textcolor{black}{P.T.}} (m-1-2)
   	(m-1-1) edge [->,very thick,blue] node [below] {\textcolor{black}{$X(x,t)$, $\tau(t)$, $\psi=A(x,t)\Psi$}} (m-1-2)
   	(m-2-1) edge [->,very thick,blue] node [below] {\textcolor{black}{P.T.}} (m-2-2);
\end{tikzpicture}
\caption{\footnotesize Connection between the stationary and parametric oscillators through the point transformation (P.T. for short). The orientation of the blue (horizontal) arrows may be inverted with the construction of the inverse point transformation. Thus, the diagram is commutative.}
\label{fig:DIA}
\end{figure}

\subsection{Linear superpositions and representation space}
\label{Seclin}

Consider the normalized superposition
\be
\vert \chi;t\rangle_{I}=\sum_{n=0}^{\infty} c_{n}\vert \varphi_{n}(t)\rangle, \quad \mbox{with} \quad \sum_{n=0}^{\infty}\vert c_{n}\vert^{2}=1, \quad c_{n} \in\mathbb{C}.
\label{eq:INV4}
\ee
We say that any regular solution of the Schr\"odinger equation~\eqref{eq:INT4}, in free-representation form, can be written as
\begin{equation}
\vert \chi;t\rangle=e^{-i\hat{I}_{2}(t) \tau (t)/\hbar}\vert\chi;t\rangle_{I}=\sum_{n=0}^{\infty}c_{n}e^{-i w(n+1/2) \tau (t)}\vert\varphi_{n}(t)\rangle=\sum_{n=0}^{\infty}c_{n}\vert \psi_{n}(t)\rangle \, .
\label{eq:INV5}
\end{equation}
Additionally, we can construct linear operators $\hat{\mathcal{O}}(t,t')$ that map elements of $\mathcal{H}(t')$ into elements of $\mathcal{H}(t)$. Using the Hubbard representation~\cite{Enr13} we may write
\be
\hat{\mathcal{O}}(t,t'):=\sum_{n,m=0}^{\infty}\mathcal{O}_{n,m}\vert\psi_{n}(t)\rangle\langle\psi_{m}(t')\vert \, , \quad \mathcal{O}_{n,m}=\langle\psi_{n}(t)\vert\hat{\mathcal{O}}(t,t')\vert\psi_{m}(t')\rangle \, ,
\label{eq:INV6}
\ee
where the coefficient $\mathcal{O}_{n,m}$ does not depend on time. In particular, for equal times $\hat{\mathcal{O}}(t):=\hat{\mathcal{O}}(t,t)$, we can construct a representation of the identity operator in $\mathcal{H}(t)$ as
\be
\mathbb{I}(t):=\sum_{n=0}^{\infty}\vert\varphi_{n}(t)\rangle\langle\varphi_{n}(t)\vert.
\label{eq:INV7}
\ee
The time-evolution operator $U(t, t')$ is obtained from~\eqref{eq:INV6} by fixing $\mathcal{O}_{n,m}=1$ for any $n,m$. From the orthogonality of the eigenfunctions at a fixed time~\eqref{eq:INN4} it follows that the action of $U(t,t')$ on any superposition~\eqref{eq:INV4} defined in $t'$ produces
\be
U(t,t')\vert\chi;t'\rangle=\sum_{n=0}^{\infty}c_{n}U(t,t')\vert\psi_{n}(t')\rangle=\sum_{n=0}^{\infty} c_{n}\vert\psi_{n}(t)\rangle=\vert\chi;t\rangle.
\ee
In turn, the time-propagator 
\be
G(x,t;x't')=\sum_{n=0}^{\infty}\psi_{n}(x,t)\psi^{*}_{n}(x',t')
\ee
is such that  
\be
\psi_{\chi}(x,t)=\langle x \vert\chi;t\rangle=\int_{-\infty}^{\infty}dx' \, G(x,t;x',t')\psi_{\chi}(x',t').
\ee
The time-propagator can be explicitly computed by using the solutions~\eqref{eq:INN3} and the summation identities of the Hermite polynomials~\cite{Olv10}. However, such a derivation is not necessary in the present work. A discussion on the matter has been recently carried out for a similar problem in~\cite{Dod73}.

\section{Coherent states}
\label{Seccs}

The simplest form to define the coherent states is to say that they ``are superpositions of basis elements to which some specific properties are requested on demand'' \cite{Ros19}. In this sense the discussion of Section~\ref{Seclin} is relevant since the capability of summing up an orthonormal set of the parametric oscillator states facilitates the construction of the corresponding (generalized) coherent states. Additionally, as the set $\{a_{2}(t),a_{2}^{\dagger}(t),\mathbb{I}(t) \}$ generates the Heisenberg Lie algebra (\ref{algebra}), one may use the conventional disentangling formulae to construct the appropriate displacement operator $\hat D(\alpha;t)$. The relevant point here is that the set $\{a_{2}(t),a_{2}^{\dagger}(t),\mathbb{I}(t) \}$, together with the invariant $\hat I$, close the oscillator algebra (\ref{eq:ALG4}). Thus, the coherent states so constructed are linear superpositions of the eigenstates of $\hat I$ which, in turn, is factorized by the time-dependent ladder operators (\ref{factor}). The resemblance of the mathematical background of the parametric oscillator to that of the stationary oscillator is, in this form, extended to the related coherent states.

Using the conventional disentangling formulae, see e.g. \cite{Ros19,Gil74}, using $a_{2}(t)$ and $a_{2}^{\dagger}(t)$, one obtains the operator
\be
\hat{D} (\alpha;t)=e^{\frac{1}{\hbar w}\left( \alpha \hat{a}_{2}^{\dagger}(t) - \alpha^{*}\hat{a}_{2}(t) \right)}=e^{-\frac{\vert\alpha\vert^{2}}{2\hbar w}}e^{\frac{\alpha}{\hbar w}\hat{a}_{2}^{\dagger}(t)}e^{-\frac{\alpha^{*}}{\hbar w}\hat{a}_2(t)}, \quad \alpha\in\mathbb{C},
\label{eq:CS1}
\ee
which produces displacements on the time-dependent ladder operators
\be
\hat{D}^{\dagger}(\alpha;t)\hat{a}_{2}(t)\hat{D} (\alpha;t)=\hat{a}_{2}(t)+\alpha, \quad \hat{D}^{\dagger} (\alpha;t)\hat{a}^{\dagger}_{2}(t)\hat{D} (\alpha;t)=\hat{a}^{\dagger}_{2}(t)+\alpha^{*}.
\label{eq:CS2}
\ee
In the Perelomov picture \cite{Per86} the coherent states $\vert\alpha;t\rangle_{I}$ are constructed by the action of $D(\alpha;t)$ on the fiducial state $\vert \varphi_{0}(t)\rangle$. From~\eqref{eq:CS2}, we find that the result 
\be
\vert \alpha; t\rangle=e^{-iw \tau (t)/2}e^{-\frac{\vert\alpha\vert^2}{2\hbar w}}\sum_{n=0}^{\infty}\left( \frac{\alpha e^{-i w \tau(t)}}{\sqrt{\hbar w}} \right)^{n} \frac{1}{\sqrt{n!}} \vert \varphi_{n}(t)\rangle,
\label{eq:CS4}
\ee
is equivalent to the one obtained in  the  Barut-Girardello picture \cite{Bar71}, where the following equation holds
\be
\hat{a}_{2}(t)\vert\alpha;t\rangle=\alpha e^{-iw \tau(t)}\vert\alpha;t\rangle.
\label{eq:CS4-1}
\ee
Although the explicit dependence on time of $\vert \alpha; t \rangle$, it is found that the related probability distribution is time-independent
\begin{equation}
\mathcal{P}_{n}(\alpha)=\vert\langle\varphi_{n}(t)\vert\alpha;t\rangle\vert^2=e^{-\frac{\vert\alpha\vert^2}{\hbar w}}\left(\frac{\vert\alpha\vert^2}{\hbar w}\right)^{n}\frac{1}{n!}.
\label{eq:CS5}
\end{equation}
Clearly, ${\cal P}_n$ is a Poisson distribution, as expected~\cite{Zel19} (compare with \cite{Una18}). In turn, the expectation values of the quadratures are as follows
\begin{subequations}
\begin{equation}
\small{\langle \hat{x} \rangle_{t}=\sqrt{\frac{2}{m}}\frac{\sigma}{w}\operatorname{Re}\alpha e^{-iwT(t)}-\gamma=\sqrt{\frac{2\vert\alpha\vert^2}{mw^{2}c}}\left[\left( \frac{w}{W_{0}}\cos\theta_{\alpha}+\frac{b}{2}\sin\theta_{\alpha} \right)q_{1} + c\sin\theta_{\alpha}q_{2} \right]-\gamma} \, ,
\label{eq:CS6-1}
\end{equation}
\begin{equation}
\langle \hat{p} \rangle_{t} = m\frac{d}{dt}\langle \hat{x} \rangle(t) = \sqrt{2m}\left(\frac{\dot{\sigma}}{w}\operatorname{Re}\alpha e^{-iwT(t)}+\frac{1}{\sigma}\operatorname{Im}\alpha e^{-iwT(t)} \right) - m\dot{\gamma} \, , 
\label{eq:CS6-2}
\end{equation}
\end{subequations}
with $\alpha=\vert\alpha\vert e^{i\theta_{\alpha}}$. If $F(t)=\gamma(t)=0$ then $\langle \hat{x} \rangle(t)$ becomes a linear combination of $q_{1,2}$ that matches with the classical result. As usual,  $\vert\alpha\vert$ and $\theta_{\alpha}$ play the role of the classical initial conditions of the system. For $F(t)\not=0$, the expected value becomes displaced by a quantity $\gamma$, so that it describes a classical oscillator subjected to the action of  a driving force~\eqref{eq:INT17}. In both cases the expected value of the momentum~\eqref{eq:CS6-2} is in agreement with the Ehrenfest theorem\cite{Sch02}, which is a property of the quadratic Hamiltonians.

On the other hand, the Heisenberg uncertainty relation is given by
\be
\left( \Delta \hat{x} \right)_{t}^{2}\left( \Delta p \right)_{t}^{2}=\frac{\hbar^{2}}{4}+\frac{\hbar^{2}}{4}\frac{\sigma^{2}\dot{\sigma}^2}{ w^{2}},
\label{eq:CS10}
\ee
with
\be
\left( \Delta \hat{x} \right)_{t}^{2}=\frac{\hbar}{2mw}\sigma^{2}, \quad \left( \Delta \hat{p} \right)_{t}^{2}=\frac{\hbar mw}{2}\left(\frac{\dot{\sigma}^{2}}{w^{2}}+\frac{1}{\sigma^2} \right).
\ee
Thus, the product (\ref{eq:CS10}) is minimized for $\dot{\sigma}=0$. The latter means that $\Delta \hat{x}$ and $\Delta \hat{p}$ are inversely proportional, up to the constant $\sfrac{\hbar}{2}$, just as this occurs in the stationary case. In the trivial situation where $\sigma \neq \sigma (t)$, from \eqref{eq:INT17} we realize that the unique solution is obtained for the constant frequency $\Omega =w^{2}/\sigma^{4} \neq \Omega(t)$, which reproduces the conventional results of the stationary oscillator. For arbitrary time-dependent $\sigma$-functions the uncertainty $\Delta \hat{x} \Delta \hat{p} \geq \sfrac{\hbar}{2}$ is  minimized at the times $t_k$ such that $\dot{\sigma}(t_k)=0$, see Section~\ref{examples} for details.

Paying attention to the product (\ref{eq:CS10}) it is clear that the variances minimize the  Schr\"odinger-Robertson inequality at any time, it is given by \cite{Rob29,Nie93,Tri94}:
\be
(\Delta \hat{x})^2(\Delta \hat{p})^2\geq\frac{\hbar^{2}}{4}+\sigma_{\hat{x},\hat{p}}^{2}, \quad \sigma_{\hat{x},\hat{p}}=\frac{1}{2}\langle \hat{x}\hat{p}+\hat{p} \hat{x} \rangle - \langle \hat{x} \rangle\langle \hat{p} \rangle,
\label{eq:SRU1}
\ee
where $\sigma_{\hat{x},\hat{p}}$ stands for the covariance function. In our case
\be
\sigma_{\hat{x},\hat{p}}=\frac{\hbar}{2}\frac{\sigma\dot{\sigma}}{w} \, .
\label{eq:SRU2}
\ee
As we can see, the coherent states of the parametric oscillator satisfy almost all the properties of the Glauber coherent states. The unique exception is that they minimize the Schr\"odinger-Robertson inequality rather than the Heisenberg uncertainty.
 
 For completeness, the coordinate representation of the coherent states is given by the wavepacket
\begin{multline}
\psi(\alpha;x,t)=  \sqrt{\frac{1}{\sqrt{2\pi}(\Delta x)_{t}}} \, \exp\left[ \frac{i}{2\hbar} \left( \int dt' F(t')\gamma(t')-\hbar w \tau (t) \right) \right] \\
\times \exp\left[ \left(-\frac{1}{4(\Delta x)^{2}_{t}}+i\frac{m}{2\hbar}\frac{\dot{\sigma}}{\sigma} \right) (x-\langle \hat{x} \rangle_{t})^{2} + \frac{i}{\hbar}\langle p \rangle_{t}x + \frac{i}{2\hbar} \langle \hat{x}\rangle_{t}\langle \hat{p}\rangle_{t} \right],
\label{eq:CS11}
\end{multline}
which is characterized by a Gaussian function with time-dependent width, the maximum of which follows  the trajectory of a classical particle under the influence of the parametric oscillator potential.

\section{Examples and discussion of results}
\label{examples}

To show the applicability of our approach we consider the results for some specific forms of the time-dependent frequency $\Omega^{2}(t)$. We take $F(t)=0$ for simplicity. With these considerations, it follows that the mapping of the position variable acquires the form
\be
X(x,t)=\frac{x+\gamma_{1} q_{1}(t)+\gamma_{2} q_{2}(t)}{\sigma(t)} \, , \quad \gamma_1,\gamma_2\in\mathbb{R} \, .
\label{eq:FP}
\ee

\subsection{$\Omega^{2}(t)=0$.}

Despite its simplicity, the null frequency $\Omega=0$ provides a connection between the solutions of the harmonic oscillator  and the free-particle systems, see e.g. \cite{Mil81,Blu96}. It is straightforward to obtain the function 
\be
\sigma(t)=\left(a+ct^2+2\sqrt{ac-w^{2}} \, t\right)^{1/2}, \quad \gamma(t)=\gamma_{1}+\gamma_{2}t ,
\label{eq:NF0}
\ee
where $a,c>0$ and $ac>w^{2}$. Then, the relation between the time parameters is given by
\be
\tau (t)=\frac{1}{w}\arctan\left[\frac{1}{w}\left(\sqrt{ac-w^{2}}+ct \right) \right] \, ,
\label{eq:NF1}
\ee
while the spatial coordinates are related through Eq.~\eqref{eq:FP}. Now, from \eqref{eq:INN1} with $a=c=w=1$, we arrive at the equivalent result
\be
\psi(x,t)=e^{i\frac{m}{\hbar}\left(\frac{tx^{2}}{1+t^2}\right)}\left(1+t^2\right)^{-1/4}\Psi\left(\frac{x}{\sqrt{1+t^2}},\arctan t \right),
\label{eq:NF2}
\ee
which has been already reported in \cite{Mil81}, p.~83. The above procedure permits the construction of coherent states for the free-particle system by means of a simple mapping of the Glauber states to the appropriate basis (similar results can be found in \cite{Bag14}). In such case, the function $\sigma$ is proportional to the width of the wave-packet which, from~\eqref{eq:NF0}, is an increasing function in time. In other words, the coherent states of a free-particle are less localized as the time goes pass.

\subsection{$\Omega^{2}(t)=\Omega_{0}^{2}>0$.}

In this case the Hamiltonian (\ref{eq:PMO1}) is of the form 
\be
\left. \hat{H} (t) \right\vert_{\Omega(t)=\Omega_{0}} =\frac{\hat{p}^{2}}{2m}+\frac{m\Omega_{0}^{2}}{2}\hat{x}^{2} \equiv \hat{H}_{osc}.
\label{eq:CF0}
\ee
That is, $\hat H(t)$ represents a stationary oscillator of frequency $\Omega_{0}$. With the pair of linearly independent functions, $q_{1}(t)=\cos(\Omega_{0} t)$ and $q_{2}(t)=\sin(\Omega_{0} t)$, the functions $\sigma$ and $\gamma$ take the form
\be
\begin{aligned}
& \sigma^{2}(t)=a\cos^{2}(\Omega_{0}t)+c\sin^{2}(\Omega_{0} t)+\sqrt{ac-\frac{w^{2}}{\Omega^{2}_{0}}}\,\sin(2\Omega_{0}t), \\[1ex]
& \gamma(t)=\gamma_{1}\cos\Omega t+\gamma_{2}\sin\Omega t.
\end{aligned}
\label{eq:CF1}
\ee
From~\eqref{eq:INV2} and~\eqref{eq:CF1} we realize that $\hat{I}(t)$ still is a time-dependent operator, which is also an invariant of the system. Consequently, the functions $\varphi_{n}(x,t)$ are not eigenfunctions of $\hat{H}$, although, they are solutions of the corresponding Schr\"odinger equation. In the special case $a=c=w/\Omega$ we obtain $\sigma(t)=w/\Omega$. In addition, for $\gamma_{1,2}\not=0$ we recover the displaced number states discussed in~\cite{Nie97} and \cite{Phi14}. For $\gamma_{1,2}=0$, the eigenfunctions $\varphi_{n}$ are simply reduced to the solutions of the stationary oscillator of frequency $\Omega_{0}$.

\subsection{$\Omega^{2}(t)=\Omega_1+\Omega_2 \tanh(k t)$.}

For $\Omega_{1}>\Omega_{2}$ the frequency $\Omega(t)$ changes smoothly from $\Omega_{1}-\Omega_{2}$ to $\Omega_{1}+\Omega_{2}$. In the limit $k\rightarrow\infty$, the function $\Omega(t)$ converges to the Heaviside step distribution $\Theta (t)$ \cite{Olv10}. In general, we have the linearly independent functions
\be
\begin{aligned}
& \widetilde{q}_{1}(t)=(1-z)^{-\frac{i}{2}g_{+}}(1+z)^{-\frac{i}{2}g_{-}} \, {}_{2}F_{1}\left( \left. \begin{aligned} -i \mu \, , \, 1-i \mu \\ 1-ig_{+}(t) \hspace{5mm} \end{aligned} \right\vert \frac{1-z}{2} \right) , \\[1ex]
& \widetilde{q}_{2}(t)=(1-z)^{+\frac{i}{2}g_{+}}(1+z)^{+\frac{i}{2}g_{-}} \, {}_{2}F_{1}\left( \left. \begin{aligned} i \mu \, , \, 1+i \mu \\ 1+ig_{+}(t) \hspace{5mm} \end{aligned} \right\vert \frac{1-z}{2} \right), \\[1ex]
& g_{\pm}=\mu \pm\frac{\Omega_2}{2k^{2}\mu}, \quad \mu=\frac{1}{k}\sqrt{\frac{\Omega_1+\sqrt{\Omega_1^{2}-\Omega_2^{2}}}{2}}, \quad z=\tanh(k t) ,
\end{aligned}
\label{eq:TDF2}
\ee
where ${}_{2}F_{1}(a,b;c;z)$ stands for the hypergeometric function \cite{Olv10}. From~\eqref{eq:TDF2} it is clear that both $\widetilde{q}_{1,2}$ are complex-valued functions. Moreover, as $\widetilde{q}_{2}(t)=\widetilde{q}^{*}_{1}(t)$, the Wronskian is the pure imaginary number $W_{r}(\widetilde{q}_1,\widetilde{q}_2)=-2ikg_{+}$. 

\begin{figure}[htb]
\centering
\includegraphics[width=0.3\textwidth]{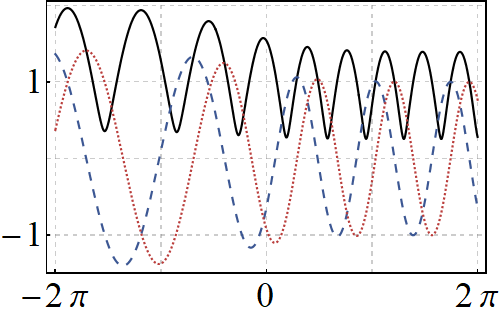}
\caption{\footnotesize The solution of the Ermakov equation~\eqref{eq:OSC7} (solid-black)  is compared  with $q_{1}(t)$ (dashed-blue) and $q_{2}(t)$ (dotted-red). In all cases the time-dependence is dictated by the frequency function $\Omega^{2}(t)=\Omega_1+\Omega_2 \tanh(kt)$, with $k=1/2$, $\Omega_1=5$, $\Omega_2=3$, and $a=c=1$.}
\label{fig:F1}
\end{figure}

Following the discussion of Appendix~\ref{ApC} we set $q_{1}=\operatorname{Re}[q_{1}]$ and $q_{2}=\operatorname{Im}[q_{1}]$ as the pair of linearly independent real solutions that are required in our approach. Then $W_{0}=kg_{+}$, and
\be
\sigma^{2}(t)=a\operatorname{Re}[q_{1}]^2+c\operatorname{Im}[q_{1}]^2+2\sqrt{ac-\frac{w^{2}}{k^{2}g_{+}^{2}}} \, \operatorname{Re}[q_{1}] \operatorname{Im}[q_{1}],
\label{eq:TDF3}
\ee
where $a,c>0$ to obtain a nodeless real-valued solution. It is worth to remember that any linear combination of Re$[q_{1}]$ and Im$[q_{1}]$ can be used to describe the classical motion of a particle under the influence of the parametric oscillator. Whereas for the quantum case the nonlinear combination~\eqref{eq:TDF3} is necessary to make any prediction. The behavior of Re[$q_{1}$], Im[$q_{1}$], and $\sigma$ is depicted in Figure~\ref{fig:F1}. It can be appreciated that the classical solutions transit from lower ($t<0$) to higher ($t>0$) frequency oscillations, as expected. The time rate of such transition is controlled by the parameter $k$. The oscillations are not exactly periodic, but they can be cosidered periodic at large enough times.

\begin{figure}[htb]

\centering
\subfigure[~$n=0$  ]{\includegraphics[width=0.3\textwidth]{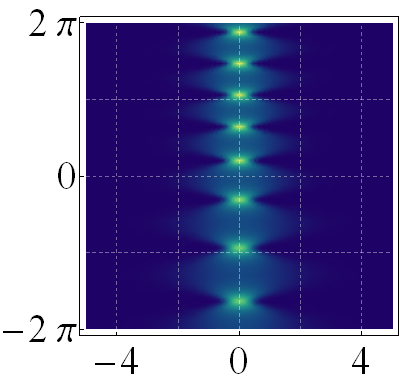} } 
\hskip1ex
\subfigure[~$n=1$ ]{\includegraphics[width=0.3\textwidth]{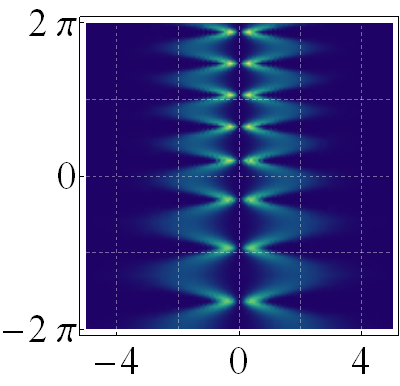} } 
\hskip1ex
\subfigure[~$n=2$  ]{\includegraphics[width=0.3\textwidth]{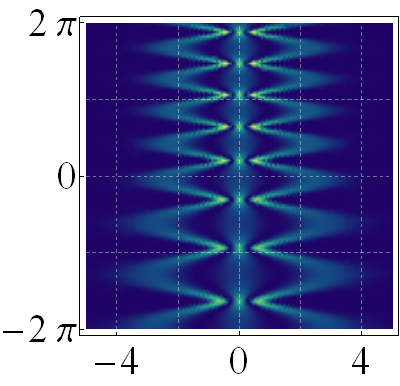} }

\caption{\footnotesize 
Probability density $\vert\varphi_{n}\vert^{2}=\vert\psi_{n}\vert^{2}$ for the indicated values of $n$ with $k=1/2,\Omega_{1}=5,\Omega_{2}=3,a=c=w=1$. The horizontal and vertical axes correspond to position and time, respectively.
}
\label{fig:F2}
\end{figure}

The probability densities of the eigenfunctions $\varphi_{n}(x,t)$ are shown in Figure~\ref{fig:F2} for $n=0,1,2$. We can appreciate that $\varphi_{0}(x,t)$ is a localized wave-packet that spreads out during a finite interval of time, then it is squeezed up to it recovers its initial configuration. Such an oscillatory property is relevant in the paraxial approximation of electromagnetic signals, for it is associated with self-focusing beams in varying media \cite{Cru17,Gre17,Gre19,Raz19}. For higher eigenfunctions there is a definite number of nodes, the position of which varies in time. Moreover, from the polynomial behavior of the solutions, it is clear that the oscillation theorem holds at each time, leading to a complete set of solutions which form a basis. The latter generates a vector space which turns out to be dynamical \cite{Ali18}.

\begin{figure}[htb]

\centering
\subfigure[]{\includegraphics[width=0.3\textwidth]{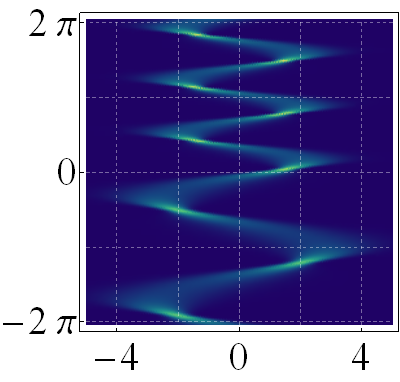} } 
\hskip1cm
\subfigure[]{\includegraphics[width=0.41\textwidth]{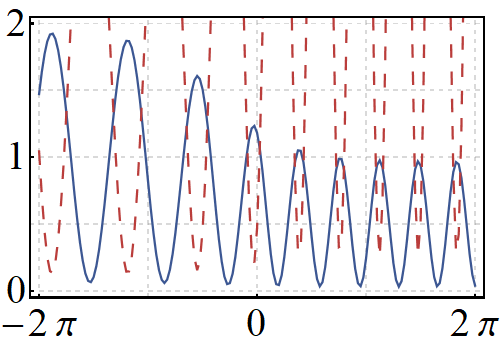} } 

\caption{\footnotesize 
(a) Probability density $\vert\psi(\alpha;x,t)\vert^{2}$ for the coherent states with $k=1/2,\Omega_{1}=5,\Omega_{2}=3,a=c=w=1$. The horizontal and vertical axes correspond to position and time, respectively. (b) Variances of the physical position $(\Delta\hat{x})^{2}_{t}$ (solid-blue) and momentum $(\Delta\hat{p})^{2}_{t}$ (dashed-red), with the same parameters as in figure~(a).
}
\label{fig:F3}
\end{figure}

On the other hand, the behavior of the coherent states in coordinate representation~\eqref{eq:CS11} and the variances associated with it~\eqref{eq:CS10} are depicted in Figure~\ref{fig:F3}. It is clear that the maximum of $\vert\psi(\alpha;xt)\vert^{2}$ follows a classical trajectory, compare with the behavior of $q_{1}(t)$ in Fig.~\ref{fig:F1}. The variance $(\Delta\hat{x})^{2}$ squeezes in time with oscillatory profile. The squeezing increases as the time goes on. On the other hand, the variance $(\Delta\hat{p})^{2}$ spreads more strongly than  its canonical counterpart. Thus, this configuration skews in favor of the localization in position, which is the desired behavior inside ion traps, as discussed in, e.g., \cite{Gla92}.

\section{Conclusions}
\label{conclu}

We have shown that the properly chosen point transformation permits to solve the Schr\"odinger equation for a wide diversity of nonstationary oscillators. Our method overpasses the difficulties that arise in the conventional approaches like the absence of the observable(s) that define(s) uniquely the state of a parametric oscillator. Namely, as the related Hamiltonian is not an integral of motion, it is usual to provide an ans\"atz in order to guess the form of the related invariant. A striking feature of our method is that the integrals of motion are automatically obtained as a consequence of the transformation, with no necessity of guessing any ans\"atz. In this context, it is to be expected that our method can be applied to study the dynamics of particles in electromagnetic traps \cite{Pau90}.

Other difficulty which is automatically fixed by our approach concerns the orthogonality of the solutions of the nonstationary oscillators. That is, in contrast with the stationary case, solving the Schr\"odinger equation for a nonstationary system, the orthogonality of the solutions is not automatically granted. We demonstrated that the orthonormality of the states of the parametric oscillator is granted by the point transformation of the states of the stationary case. The dynamical algebra, in turn, is also inherited from the stationary oscillator algebra. The latter results laid the groundwork to construct the corresponding coherent states, which inherit all the properties of the Glauber states with the exception that they minimize the Schr\"odinger-Robertson inequality rather than the Heisenberg uncertainty. 

Additional applications may include the propagation of electromagnetic signals in waveguides, where the Helmholtz equation is formally paired with the Schrödinger one \cite{Man08,CruT15a,CruT15b}, and the self-focusing is relevant \cite{Cru17,Gre17,Gre19,Raz19}. Finally, the approach can be extended to study supersymmetric structures in quantum mechanics \cite{Mie04} with time-dependent potentials \cite{Zel17a,Con17}

\appendix
\section{Point transformation}
\label{ApA}

\renewcommand{\thesection}{A-\arabic{section}}
\setcounter{section}{0}  

\renewcommand{\theequation}{A-\arabic{equation}}
\setcounter{equation}{0}  

The detailed derivation of Equations~(\ref{eq:INT11})-(\ref{eq:INT12}) in terms of point transformations \cite{Ste93} is as follows. We first consider the explicit dependence of $X$, $\tau$, and $\psi$ on the set $\{x,t; \psi(x,t) \}$ given in (\ref{eq:INT5})-(\ref{eq:INT5-1}). The mapping from $S_{in}$ to $S_{def}$, see Eq.~(\ref{eq:INT6-1}), must be such that nonlinearities are not present in $S_{def}$. In general, it is expected to find
\be
\Psi_{\tau} =G_{1}\left(x,t;\psi, \psi_t,\psi_x \right), \quad \Psi_{X,X} =G_{2}\left(x,t;\psi, \psi_t, \psi_x, \psi_{x,x} \right).
\label{eq:INT6}
\ee
Using \eqref{eq:INT5-1} and \eqref{eq:INT6}, the Schr\"odinger equation of the stationary oscillator \eqref{eq:INT1} becomes a partial differential equation of the desired form $S_{def}$. To be concrete, we have
\be
\frac{d\Psi}{dx}= \Psi_X X_x + \Psi_{\tau} \tau_x, \quad 
\frac{d\Psi}{dt}= \Psi_X X_t + \Psi_{\tau} \tau_t.
\label{eq:INT7}
\ee
Equivalently, from  (\ref{eq:INT5-1}) one gets
\be
\frac{d\Psi}{dx} = G_{\psi} \psi_x + G_x, \quad
\frac{d\Psi}{dt} = G_{\psi} \psi_t + G_t.
\label{G}
\ee
The system \eqref{eq:INT7}-\eqref{G} includes $\Psi_X$ and $\Psi_{\tau}$ as unknown functions, the solutions of which are
\be
\begin{aligned}
& \Psi_ X = \frac{1}{J(x,t)}\left( \tau_{t}G_{\psi} \psi_x - \tau_{x}G_{\psi} \psi_t + \tau_{t}G_{x} - \tau_{x}G_{t} \right), \\[0.5ex]
& \Psi_{\tau}=\frac{1}{J(x,t)}\left( -X_{t}G_{\psi} \psi_x +X_{x}G_{\psi} \psi_t - X_{t}G_{x} + X_{x}G_{t} \right),
\end{aligned}
\label{eq:INT8}
\ee
where $J(x,t)=X_{x}\tau_{t}-X_{t}\tau_{x}\not=0$ stands for the Jacobian of the transformation. In similar form
\be
\frac{d^2 \Psi}{dx^2} = \Psi_{X,X} X_x^2 + \Psi_{\tau,\tau} \tau_x^2 + 2 \Psi_{X,\tau} X_x \tau_x + \Psi_X X_{x,x} + \Psi_{\tau} \tau_{x,x},
\label{eq:INT9}
\ee
 equivalently
\be
\frac{d^2 \Psi}{dx^2} = G_{\psi} \psi_{x,x} + 2 G_{x,\psi} \psi_x + G_{\psi,\psi} \psi_x^2 + G_{x,x}.
\label{G2}
\ee
To simplify the calculations, with no loss of generality, we take a function $\tau(x,t)$ that depends on the time parameter $t$ only, $\tau=\tau(t)$. The Jacobian  is immediately simplified 
\be
J=J(x,t) = X_{x}\tau_{t}.
\label{Jac}
\ee
On the other hand, the function $G_{\psi,\psi}$ produces the nonlinearity $\psi_x^2$ in (\ref{G2}) that is not present in $S_{def}$. Therefore we must impose the condition $G_{\psi,\psi}=0$, which permits to factorize the function $\Psi$ in \eqref{eq:INT5-1} as follows
\be
\Psi = G (x,t;\psi(x,t)) =A(x,t)\psi(x,t),
\label{eq:INT9-1}
\ee
with $A(x,t)$ a complex-valued function to be determined. Therefore, from \eqref{eq:INT8} and~\eqref{eq:INT9} we arrive at the expressions
\be
\begin{aligned}
& \Psi_{\tau} = \frac{X_{x}}{J}\left[ - A \frac{X_{t}}{X_{x}} \psi_x +  A \psi_t+\left( A_{t}-\frac{X_{t}}{X_{x}} A_{x}\right) \psi \right], \\[0.5ex]
& \Psi_X = \frac{\tau_{t}}{J}\left[A \psi_x + A_{x}\psi \right], \\[0.5ex]
& \Psi_{X,X} = \frac{1}{X_{x}^{2}}\left[ A \psi_{x,x}+\left( 2A_{x} - A \frac{X_{xx}\tau_{t}}{J} \right) \psi_x + \left(A_{xx}-\frac{X_{xx} \tau_{t}}{J}A_{x} \right)\psi \right].
\end{aligned}
\label{eq:INT10}
\ee
After substituting Eqs.~\eqref{eq:INT9}-\eqref{eq:INT10} in \eqref{eq:INT1}, together with some arrangements, we finally have
\[
i\hbar \psi_t+\frac{\hbar^{2}}{2m}\frac{\tau_{t}}{X_{x}^{2}} \psi_{x,x} + B(x,t) \psi_x - V(x,t)\psi = 0,
\]
where
\[
\begin{aligned}
& B(x,t)=-i\hbar\frac{X_{t}}{X_{x}}+\frac{\hbar^{2}}{2m}\frac{\tau_{t}}{X_{x}^{2}}\left( 2\frac{A_{x}}{A}-\frac{X_{xx}}{X_{x}} \right) ,\\[1ex]
& V(x,t)=-i\hbar\left(\frac{A_{t}}{A}-\frac
{X_{t}}{X_{x}}\frac{A_{x}}{A} \right)-\frac{\hbar^{2}}{2m}\frac{\tau_{t}}{X_{x}^{2}}\left( \frac{A_{xx}}{A}-\frac{X_{xx}}{X_{x}}\frac{A_{x}}{A} \right)+\frac{\tau_{t}}{2}m w^{2}X^{2}(x,t).
\end{aligned}
\]

\appendix
\setcounter{section}{1}  
\section{Zero point energy term}
\label{ApB}

\renewcommand{\thesection}{B-\arabic{section}}
\setcounter{section}{0}  

\renewcommand{\theequation}{B-\arabic{equation}}
\setcounter{equation}{0}  

Consider the Schr\"odinger equations
\be
i\dot{\Phi}=-\frac{\partial^2}{\partial x^2}\Phi + \widetilde V(x,t) \Phi, \quad \Phi=\Phi(x,t),
\label{eq:TSP1}
\ee
and
\be
i\dot{\Psi}=-\frac{\partial^2}{\partial x^2}\Psi + V(x,t)\Psi, \quad \Psi=\Psi(x,t),
\label{eq:TSP2}
\ee
with $\widetilde V(x,t) =V(x,t)+V_{0}(t)$. Using $\Phi(x,t)=h(t)\Psi(x,t)$ in \eqref{eq:TSP1} we arrive at a differential equation for $h(t)$, the solution of which produces
\be
\Phi(x,t)= \exp \left[ {-i\int^{t} dt' \, V_{0}(t')} \right] \Psi(x,t).
\label{eq:}
\ee
That is, if $\widetilde V(x,t)$ differs from $V(x,t)$ by an additive time-dependent term $V_0(t)$, the solutions of (\ref{eq:TSP1}) and (\ref{eq:TSP2}) coincide up to a global phase that depends on time. Of course, if $V_0 \neq V_0(t)$, then $\Phi(x,t)$ and $\Psi(x,t)$ belong to the same equivalence class (ray) in the space of states. 

\appendix
\setcounter{section}{2}  
\section{The Ermakov equation}
\label{ApC}

\renewcommand{\thesection}{C-\arabic{section}}
\setcounter{section}{0}  

\renewcommand{\theequation}{C-\arabic{equation}}
\setcounter{equation}{0}  

The Ermakov equation \cite{Erm08}
\be
\ddot{\sigma}+\Omega^{2}(t)\sigma=\frac{w^2}{\sigma^{3}}, \quad w>0,
\label{Erma}
\ee
is well known in the literature and finds many application in physics \cite{Cas13,Sch13,Cru15,Cru16,Zel19,Sch18,Ros15,Bla18,Pad18,Gal18,Cru17,Gre17,Gre19,Raz19}. It arises quite naturally in the studies of parametric oscillators \cite{Cas13,Sch13,Cru15,Cru16,Zel19}, in the description of structured light in varying media \cite{Cru17,Gre17,Gre19,Raz19}, and in the study of non-Hermitian Hamiltonians with real spectrum \cite{Sch18,Ros15,Bla18}. The key to solve (\ref{Erma}) is to consider the homogeneous linear equation
\be
\ddot{q}+\Omega^{2}(t)\,q=0,
\label{eq:OSC6}
\ee
which coincides with the equation of motion for a classical parametric oscillator. Consider two solutions, $q_1$ and $q_2$, and the related Wronskian $W(q_1,q_2)=q_1\dot{q}_2-\dot{q}_1 q_2$.  It is straightforward to show that $W(q_1,q_2)$ is a constant in time, and different from zero if the involved solutions are linearly independent. 

Using two linearly independent solutions, $q_1$ and $q_2$, of (\ref{eq:OSC6}) we have $W(q_1,q_2)= W_0 = \mbox{const}$. Then, following \cite{Erm08}, the solution of (\ref{Erma}) is of the form
\be
\sigma (t) = [ a q_1^2(t) + b q_1(t) q_2(t) + cq_2^2(t) ]^{1/2},
\ee
where $\{a,b,c\}$ is a set of real constants. To get a function $\sigma>0$, it is necessary to impose the condition $b^2-4ac=-4\frac{w^2}{W_{0}^2}$, with nonnegative constants $\{a,b,c\}$ \cite{Ros15,Bla18}.

If, by chance, the accessible solution of (\ref{eq:OSC6}) is a complex-valued function, say $\widetilde{q}:\mathbb{R}\rightarrow\mathbb{C}$, it follows that its complex conjugated $\widetilde{q}^{*}$ is a second linear independent solution. Then, without loss of generality, the real and imaginary parts of $\widetilde{q}$ can be used as the pair of linearly independent solutions one is looking for. That is, $q_{1}=\operatorname{Re}[\widetilde{q}]$ and $q_{1}=\operatorname{Im}[\widetilde{q}]$. In this form the $\sigma$-function, as well as the Jacobian of the transformation, are well-behaved. Then, they produce singular-free transformation functions $X(x,t)$ and $\tau(x,t)$.

\section*{Acknowledgment}

This research was funded by Consejo Nacional de Ciencia y Tecnolog\'ia (Mexico), grant number A1-S-24569. K. Zelaya acknowledges the support from the Laboratory of Mathematics Physics, Centre de Recherches Math\'ematiques, through a postdoctoral fellowship.


\end{document}